# Early Detection of Tuberculosis with Machine Learning Cough Audio Analysis: Towards More Accessible Global Triaging Usage

Chandra Suda[1],

[1] Bentonville, Arkansas

## Abstract

Tuberculosis (TB), a bacterial disease mainly affecting the lungs, is the leading infectious cause of mortality worldwide. To prevent TB from spreading within the body, which causes life-threatening complications, timely and effective anti-TB treatment is crucial. Cough, an objective biomarker for TB, is a triage tool that monitors treatment response and regresses with successful therapy. Current gold standards for TB diagnosis are slow or inaccessible, especially in rural areas where TB is most prevalent. In addition, current machine learning (ML) diagnosis research, like utilizing chest radiographs, is ineffective and does not monitor treatment progression. To enable effective diagnosis, an ensemble model was developed that analyzes, using a novel ML architecture, coughs' acoustic epidemiologies from smartphones' microphones to detect TB. The architecture includes a 2D-CNN and XGBoost that was trained on 724,964 cough audio samples and demographics from 7 countries. After feature extraction (Mel-spectrograms) and data augmentation (IR-convolution), the model achieved AUROC (area under the receiving operator characteristic) of 88%, surpassing WHO's requirements for screening tests. The results are available within 15 seconds and can easily be accessible via a mobile app. This research helps to improve TB diagnosis through a promising accurate, quick, and accessible triaging tool.

## Introduction

Tuberculosis (TB), a bacterial infectious disease, is one of the top 10 causes of mortality worldwide in low-income countries and the leading cause of mortality from an infectious agent in 2022, resulting in approximately 10 million new infections and 1.4 million deaths (World Health Organization, 2022). During an initial infection, the immune system forms a granuloma to keep the bacteria from spreading. The bacteria will remain in the lungs, but the body is protected from disease by the granuloma—defined as latent TB. It is estimated that 1/5 of the world's population, about 1.6 billion people, are latently infected (Ding et al., 2022). When TB escapes from the granuloma, which can occur immediately after infection or never, and starts destroying the lungs, it is called pulmonary TB (TB Disease). Extra-pulmonary TB occurs when TB bacteria infect other parts of the body, causing potentially life-threatening complications. As such, timely treatment of anti-TB drugs is important, which can be achieved with an accurate and accessible early diagnosis. In addition, resources used to diagnose and control TB were diverted to the COVID-19 pandemic, causing a devastating impact on TB programs (Zimmer et al., 2021).

    The current reference standards for TB testing are sputum culture, smear microscopy, and GeneXpert PCR machine. However, they have limitations including expensiveness, requiring access to not widely available medical facilities, time-consuming, and inability to detect drug resistance

(Kik et al., 2014; Vongthilath-Moeung et al., 2021). Especially in rural communities, early diagnosis, and thus timely treatment, is a problem.

Deep learning techniques for TB like Computer-Aided Diagnostics (CAD) systems show early promising signs in detecting TB from posterior-anterior chest radiographs (Oloko-Oba and Viriri, 2022). But those state-of-the-art deep learning techniques are not effective in developing countries, where TB is most prevalent, because acquiring chest X-rays are not practical due to the lack of resources and skilled radiologists (Pedrazzoli et al., 2016; Pande et al., 2015). The simplest form of TB triaging relies on self-reported symptoms. Although this is a low-cost method, it has low specificity, resulting in over-testing. Existing solutions include questionnaire-based tools that have been used to evaluate the severity of coughs. For example, Leicester Cough Questionnaire (LCQ) is a validated self-completed questionnaire that measures the quality of life of people with chronic cough. LCQ was used to evaluate cohorts of people with TB undergoing anti-TB treatment (Birring et al., 2003). While they are easy to implement, they are subject to bias in self-perception, limiting their clinical application.

Current gold standard TB testing solutions are insufficient because they are resource intensive and impractical in rural areas, while the latest state-of-the-art ML solutions are neither large and diverse nor applicable through a smartphone microphone (Topol, 2020; Rutjes, 2006). In addition, the predictive ML-based monitoring of TB through coughs through the treatment cascade is lacking.

Furthermore, there currently isn't an effective method for monitoring patient anti-TB drug treatment. If there are dosage irregularities, like missing a dose in their treatment regimen, then TB may not be fully eradicated. If some of the bacteria survive and stay hidden, it can lead to TB relapse and even drug resistance.

Instead, this paper proposes a new method utilizing cough audio from a smartphone's microphone to detect whether the person has TB or not. Coughing is a common symptom of respiratory disease and is caused by an explosive expulsion of air. It has also been postulated that the glottis behaves differently under different pathological conditions, and this makes it possible to distinguish between coughs due to asthma, bronchitis and pertussis (Korpáš et al. 1996). The recent development of cough detection and recording apps provided a discrete and easy way to acquire diverse and large cough data. This model will use data from the cough recording applications—including Hyfe Research, AI4COVID-19, and ResAppDx (Imran et al., 2020; Moschovis et al., 2021). The training dataset went through similar conditions that the deployment (mobile app) will be in, allowing for reliability.

Tuberculous Drug-Induced Liver Injury (TB-DILI) is a serious and common side effect caused by hepatotoxicity from anti-TB drugs such as pyrazinamide, ethambutol, isoniazid, and rifampin (Saukkonen et al., 2006). Currently, liver tests usually include biochemical parameters of blood, such as alanine transaminase (ALT). It is difficult to distinguish DILI from non-DILI based on these indicators since test results are largely consistent (Peifang et al., 2020).

# Methodology

Datasets

Model training and testing data was obtained from the CODA TB DREAM Challenge (synapse). The full dataset encompassing 724,694 cough audio samples and 1,105 patients (Table 3) (Synapse — Sage Bionetworks, n.d.; Grants NIH, n.d.). Adults (18 years and older) with presumptive TB with greater than two weeks of cough were enrolled at participating clinics (India, Philippines, South Africa, Uganda, Vietnam, Tanzania, and Madagascar) and cough sounds recorded with a smartphone using the Hyfe Research Application ("Hyfe AI - Detect & Quantify Cough"). Participants were instructed to cough three times but additional, spontaneous, coughs were recorded as well. Standard clinical and demographic information was collected at enrollment and TB-positive participants were defined as positive on GeneXpert MTB/RIF, GeneXpert MTB/Ultra, or culture reference standards.

Audio recordings of 0.5 seconds were first classified by the Hyfe Application as cough (as opposed to other background noises) and only recordings with a prediction as cough greater than 0.85 were used.

Methodology Overview

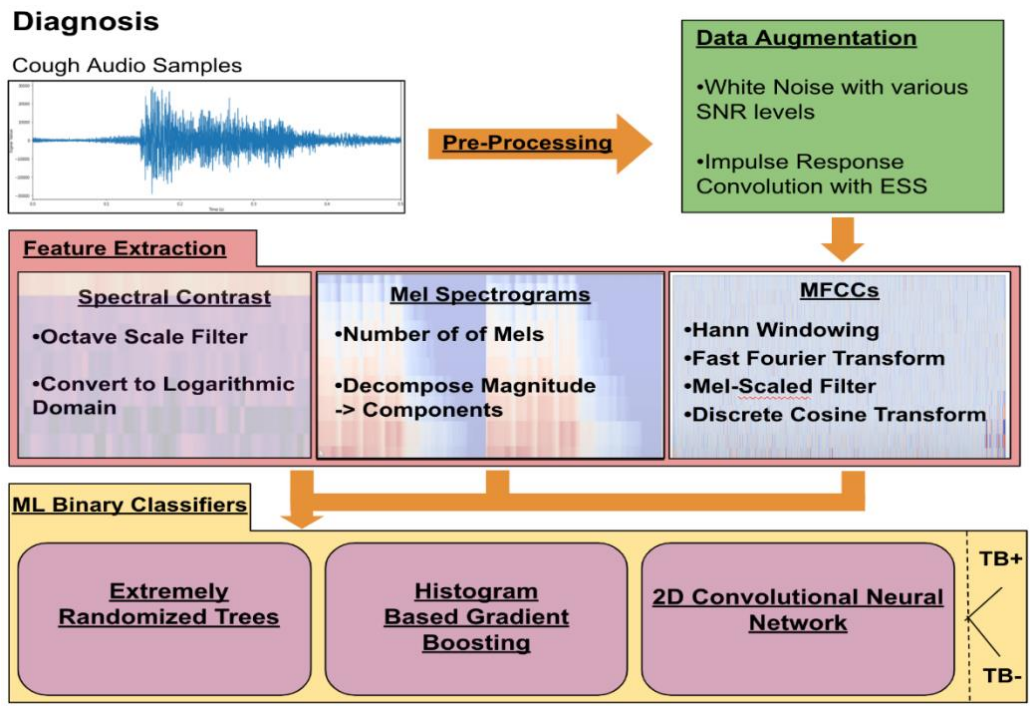

*Figure 1: Architectural pipeline overview*

Audio data has been preprocessed with the below processes. Next, the data was augmented to ensure valid real-world performance and then feature extraction was performed to train ML binary classifiers on the most valuable data.

## Preprocessing and Data Augmentation

Audio captures were converted using a sample rate of 44.1 kHz with 16 bits per sample were pre-processed using the Python libraries Pandas (Pandas, 2018), Torch (*Torch.library — PyTorch 1.12 Documentation*, n.d.), and TorchAudio (*Torch.library — PyTorch 1.12 Documentation*, n.d.). After rechanneling the audio to mono, encodings were created for the clinical and demographic dataset using the "one-hot encoding" technique. All values that were not quantitative were encoded into either 1s or 0s. For example, an 'Yes' for presenting symptoms (i.e., Weight Loss) was encoded as 1 while a 'No' was encoded as 0. This is needed as the some of the models cannot operate on label data directly (Brownlee, 2017). Variables with numerical values (i.e. Weight) were transformed using Min-Max Scaling to decrease the effects of outliers.

To increase the algorithm's robustness in real-world use cases which may have background noises or room echoes and to decrease the probability of model over fitting, data augmentation was used by artificially increasing the training sets by introducing noise into the training data. Total data augmentation was 50% of the training data (n=503063) and two types of data augmentation were employed: (1) adding white noise with varying signal-to-noise ratio (SNR) levels calculated by 10log(Signal $RMS^2$/Noise $RMS^2$) where Signal RMS is the Root Mean Squared (RMS) value of signal and Noise RMS is RMS of noise, and, (2) Impulse Response (IR) which was computed using Exponential Sine Sweep (ESS) (see Supplementary Materials). The data to augment was selected by randomizing the data. The split between ESS and SNR augmentations was 50:50 in the data. The SNR ratio was varied between 0.0:1.0 and 0.9:1.0. For the ESS, the scale was 0.5 divided by np.amax(channel_1) times np.amax(channel_2).

## Feature Extraction

Three features were extracted using the Python Librosa library (McFee et al., 2015): Mel-Frequency Cepstral Coefficients (MFCCs), Mel-Scale Spectrogram, and Spectral Contrast. MFCCs were calculated following (citation).

, after the windowing operation, the Fast Fourier Transform (FFT) was applied to find the power spectrum of each frame. Afterward, the Mel scale is used to perform filter bank processing on the power spectrum. Mel-scaled filters are calculated from physical frequency (f) by the following Figure 2. After converting the power spectrum to the logarithmic domain, discrete cosine transform (DCT) was applied to the audio signal to measure the MFCC coefficients.

$$f_{mel} = 2595\log_{10}\left(1 + \frac{f}{700}\right)$$

*Figure 2: MFCC Equation*

Mel-scaled Spectrograms were calculated following (citation) with the window size of 2048 and the hop length of 512, and Mels equal to 128, allowing an evenly spaced frequency. Finally, the magnitude of the signal was decomposed into components corresponding to the frequencies in the

Mel scale. Figure 3 showcases the FTT definition used, and Figure 4 shows an example picture of the audio converted to a Mel Spectrogram image.

$$y[k] = \sum_{n=0}^{N-1} e^{-2\pi j \frac{kn}{N}} x[n]$$

Figure 3: Definition of the FFT y[k] of length of the length-sequence x[n]

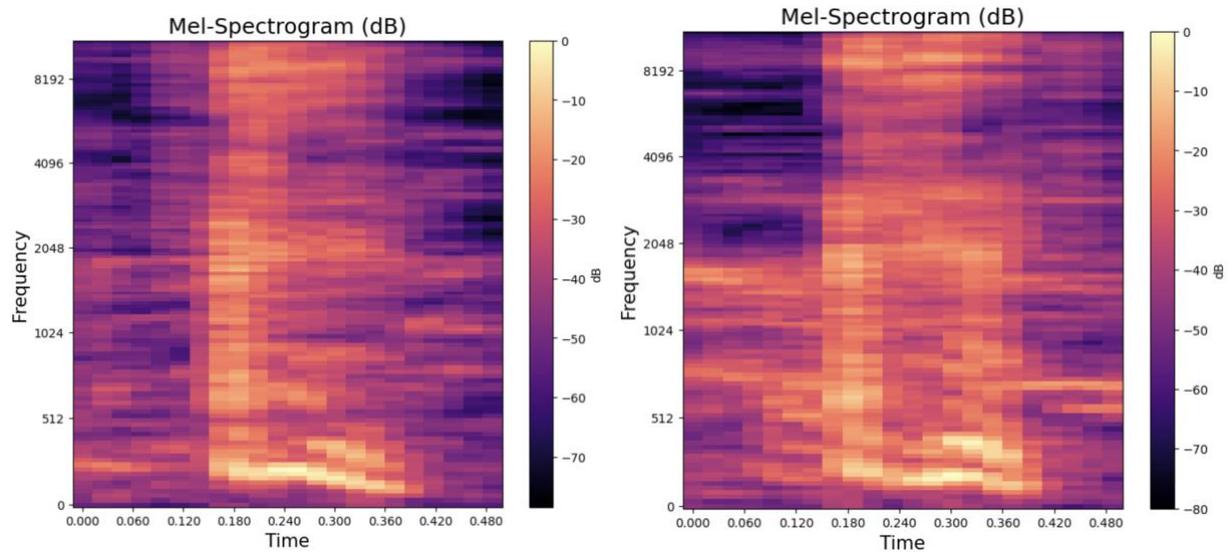

Figure 4: Example image of a TB Positive (right) and Negative (left) Mel Spectrogram

For extracting the Spectral Contrast, FTT was performed to obtain the frequency spectrum. Using eight Octave-scale filters, the frequency domain was partitioned into sub-bands. The number of frequency bands was set to be 8. The strength of spectral valleys, peaks, and their differences were evaluated in each sub-band. Mel-spectrogram is represented in image format while MFCCs and

## Machine Learning Binary Classifiers

Three ML and deep learning models were tested, but the primary three were Extremely Randomized Trees (Extra-Trees), Histogram-Based Gradient Boosting (HGBoost), and 2-Demonsional Convolutional Neural Networks (2D-CNN). Extra-Trees fit multiple random decision trees to each sub-sample of the dataset, avoiding overfitting and improving detection accuracy and have been previously demonstrated to discriminate patients with chronic obstructive pulmonary disease with an accuracy of 89% (Swaminathan et al., 2017). Gradient Boosted Random Forest has been shown to have better performance in less inference time and has been shown to diagnose COVID-19 from cough recordings with an accuracy of 87% (Chung et al., 2021). The 2D CNN is a classic, highly effective image processing ML. MFFCs and Mel-Spectrograms were extracted for the 2D-CNN to be trained on. The 2D CNN was more optimal compared to a 1D CNN because the coughs' frequencies are closer together and are semantically similar by gaining more signal from another dimension. After testing all three models and various combinations of hyperparameter tuning.

| Name | Metric | Value |
| --- | --- | --- |
| 2D-CNN | AUC | 88% |
| HGBoost | AUC | 83% |
| Extra-Trees | AUC | 76% |

*Table 1: Comparison of AUC scores for the three models*

## Classifier Evaluation

The novel 2D-CNN architecture for TB coughs takes the only Mel-Spectrogram Contrast of each audio file as the input and returns the probabilities for each class (1 and 0). The other two ML models take all three features. The architecture specifics are in Figure 5 below.

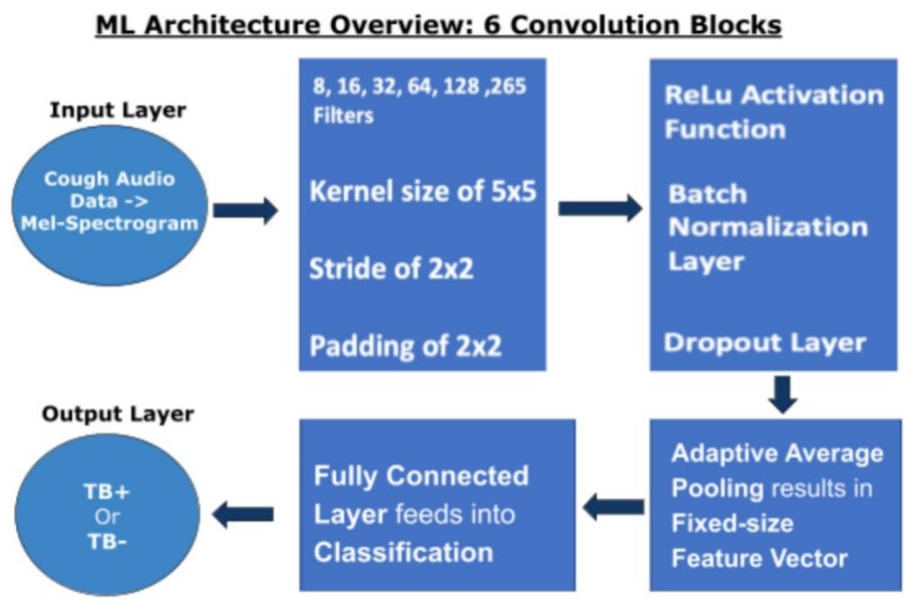

*Figure 5: 2D-CNN Architecture Layers and Parameters*

Specific training hyperparameters are listed in Table 2. The framework used for the model developed was PyTorch. The model also used the "OneCycleLR" learning rate scheduler. This specific learning rate scheduler reduces the learning rate as the epochs increase.

| Hyperparameters | Value |
| --- | --- |
| Mini Batch Size | 16 |
| Learning Rate | 6e-5 |
| Number of Epochs | 150 |
| Number of Convolutional Blocks | 6 |
| Kernal Size | 5x5 |
| Dropout Rate | 0.25 |
| Loss Function | Cross Entropy Loss |
| Optimizer | AdamW |
| Anneal Strategy | Linear |

*Table 2: 2D-Classifier Hyperparameters*

## Results

The results of the data analysis are summarized below. BMI was calculated using standard height and weight.

| Characteristic | Quantity |
|---|---|
| Audio Captures (n) | 724,964 |
| Class (Mean) | 0.61226 |
| Class (SD) | 0.48 |
| TB Positive Audio Captures (n) | 443707 (60%) |
| TB Negative Audio Captures (n) | 280986 (40%) |

*Table 3: Table showcasing properties of all audio data (longitudinal + solicited).*

|  | TB Positive | TB Negative |
|---|---|---|
| Subjects (n) | 297 | 808 |
| Age in Years: Mean + SD | 37.55 ± 14.84 | 42.06 ± 15.28 |
| Age in Years: Range | 18-83 | 18-85 |
| Sex at Birth: Male, n (%) | 195 (49%) | 393 (49%) |
| Sex at Birth: Female, n (%) | 202 (51%) | 415 (51%) |
| Initial Vital Status: Height (cm) | 163 ± 8.49 | 160.99 ± 8.79 |
| Initial Vital Status: Weight (kg) | 51.84 ± 9.24 | 59.84 ± 14.41 |
| Initial Vital Status: BMI (kg/m$^2$) | 19.3 ± 5.32 | 23.1 ± 5.54 |
| Initial Vital Status: Heart Rate (bpm) | 94.95 ± 19.61 | 82.94 ± 14.27 |
| Initial Vital Status: Temperature (C°) | 36.96 ± 0.66 | 36.64 ± 0.46 |
| History of Illness: Prior TB Exposure, n (%) | 48 (16%) | 151 (19%) |
| History of Illness: P-TB Diagnosis, n (%) | 44 (15%) | 136 (17%) |
| History of Illness: EP-TB Diagnosis, n (%) | 4 (1%) | 13 (2%) |
| Presenting Symptoms: Weight Loss, n (%) | 228 (77%) | 397 (49%) |
| Presenting Symptoms: Fever n (%) | 199 (67%) | 298 (37%) |
| Presenting Symptoms: Night Sweats, n (%) | 189 (62%) | 295 (37%) |
| Presenting Symptoms: Hemoptysis, n (%) | 64 (22%) | 84 (10%) |

| Cough Duration in Days: Mean + SD | 53.29 ± 49.51 | 44.73 ± 56.74 |

*Table 4: Table showcasing properties of tabular data. Initial is recorded at first presentation.*

## Performance Metrics

As TB+/TB- is a binary classification task, Binary Cross Entropy Loss is used as the training loss as shown in Figure 7 below.

$$L = -\frac{1}{m} \sum_{i=1}^{m} y_i \cdot \log(\hat{y}_i)$$

*Figure 7: Formula used for Binary Cross Entropy Loss*

Accuracy, true positive rate (TPR), and false positive rate (FPR) were used as the secondary endpoints for the validation (Figure 8).

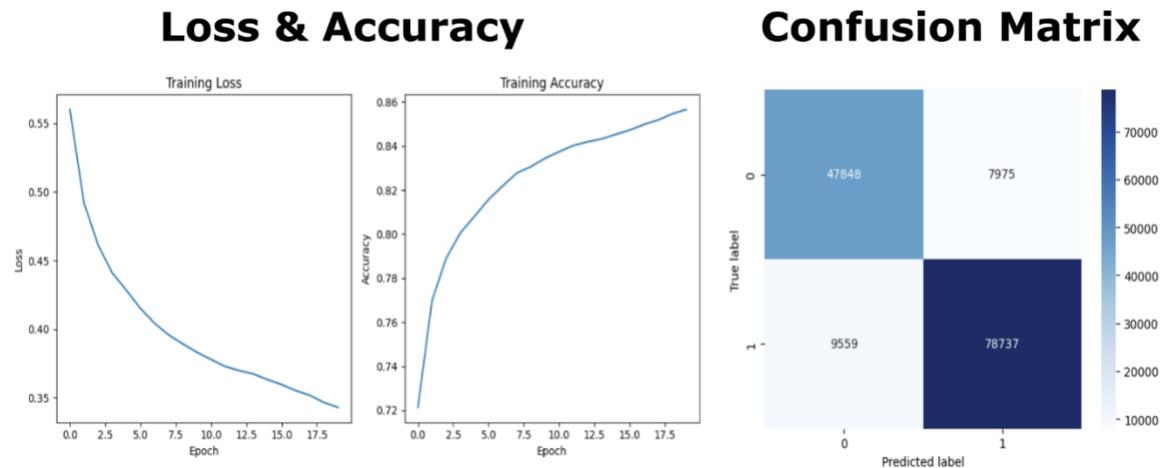

*Figure 8: Loss, Accuracy, and Confusion Matrix of results*

The primary endpoint was the area under the receiver operating characteristic (AUROC). AUROC is a metric optimal for cases when performance needs to be gauged in respect to the FPR and TPR with varying thresholds (Hajian-Tilaki, 2013). Due to the slight class imbalance of TB Positive to Negative (60-40), the FPR and TPR rate provides a less biased estimation of performance. The model achieved a high AUROC of 88% as shown in Figure 9.

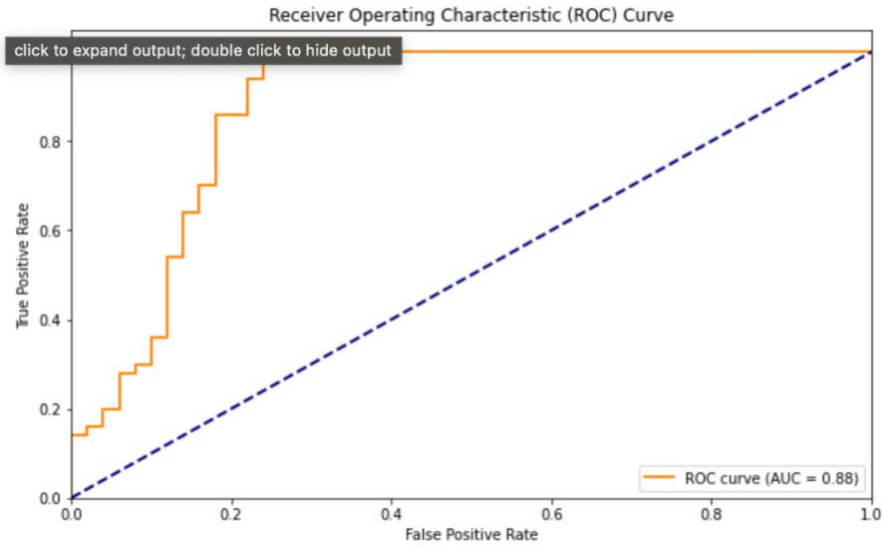

*Figure 9: ROC Curve and AUROC Score of Testing Data; TPR = Sensitivity and FPR = Specificity*

These results were then validated through a stratified 5-fold cross validation system.

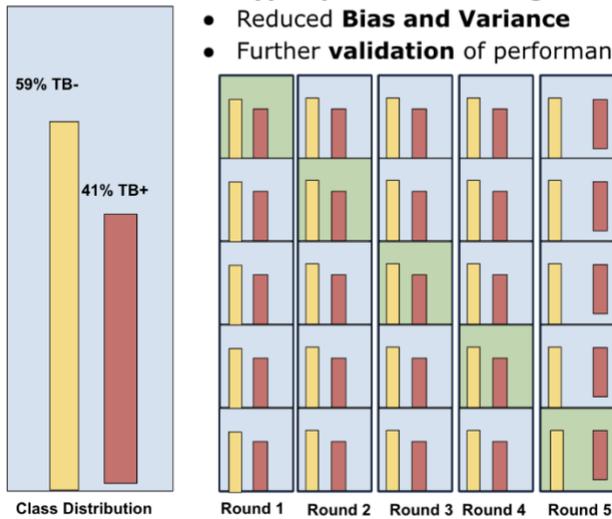

*Figure 10: Overview of stratified 5-fold cross validation*

# Conclusion and Future Work

This architecture and model indicate promise for a highly accurate triaging tool that can aid clinicals as a "first step". The ensemble model surpassed WHO requirements for a community-based triage test (90% sensitivity and 70% specificity). Early detection of TB can help increase survival outcomes greatly and help reduce spread. The accurate TB detection system from cough audio results in a highly accessible system, including future expansion to children. It is fast, low-cost, and deployable through free mobile apps. This novel framework can aid WHO's goal of eradicating TB by 2030, increasing access to quality health care in rural areas. Future research includes expansion to other cough-based diseases like COPD, Pertussis, Pulmonary embolism, etc.

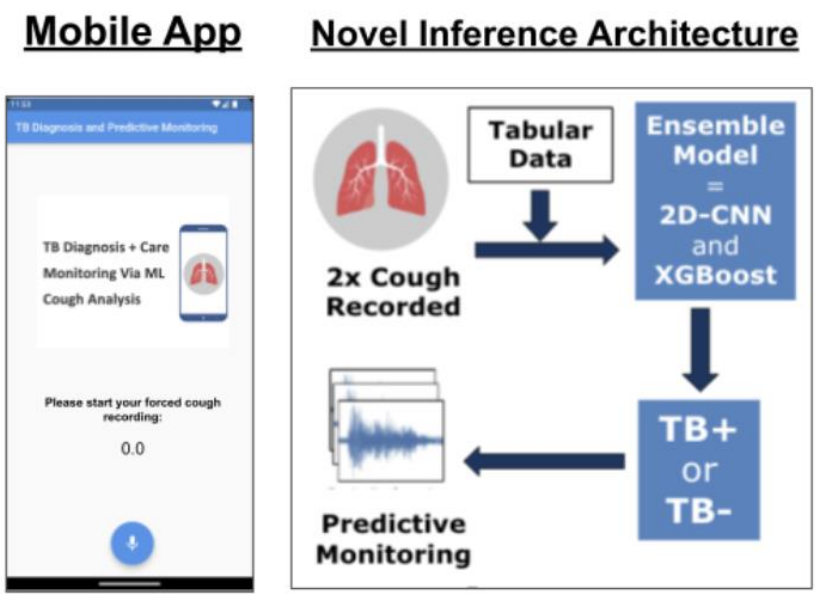

*Figure 11: Example inference architecture and mobile app*

## Acknowledgement

Challenge participants are permitted to use, publish and present the Challenge data and Challenge results, after the embargo period, provided they acknowledge the data contributors as follows: "The datasets used for the analyses described were contributed by Dr. Adithya Cattamanchi at UCSF and Dr. Simon Grandjean Lapierre at University of Montreal and were generated in collaboration with researchers at Stellenbosch University (PI Grant Theron), Walimu (PIs William Worodria and Alfred Andama); De La Salle Medical and Health Sciences Institute (PI Charles Yu), Vietnam National Tuberculosis Program (PI Nguyen Viet Nhung), Christian Medical College (PI DJ Christopher), Centre Infectiologie Charles Merieux Madagascar (PIs Mihaja Raberahona & Rivonirina Rakotoarivelo), and Ifakara Health Institute (PIs Issa Lyimo & Omar Lweno) with funding from the U.S. National Institutes of Health (U01 AI152087), The Patrick J. McGovern Foundation and Global Health Labs. They were obtained as part of the COugh Diagnostic Algorithm for Tuberculosis (CODA TB) DREAM Challenge DREAM Challenge through Synapse [**syn31472953**]."